\documentclass[review]{elsarticle}

\usepackage{lineno,hyperref}
\modulolinenumbers[5]

\usepackage{graphicx}
\usepackage{bm}
\usepackage{latexsym}
\usepackage{epsfig}
\usepackage{tikz}
\usepackage{gensymb}
\usepackage{amsmath}
\usepackage{amssymb}

\begin{document}

\begin{frontmatter}

\title{Absence of the Higuchi bound in a family of alternative linear massive spin-2 models}


\author{Hemily G.M. Fortes\corref{mycorrespondingauthor}}
\cortext[mycorrespondingauthor]{Corresponding author}
\ead{hemily.gomes@gmail.com}
\address{Divis\~ao de Astrof\'isica, Instituto Nacional de Pesquisas Espaciais, \\
Avenida dos Astronautas 1758, S\~ao Jos\'e dos Campos, SP, 12227-010, Brazil}

\author{M\'arcio E.S. Alves}
\ead{marcio.alves@unesp.br}
\address{Universidade Estadual Paulista (UNESP), Instituto de Ci\^encia e Tecnologia \\ S\~ao Jos\'e dos Campos, SP, 12247-004, Brazil}




\begin{abstract}
There is a well known result from the Fierz-Pauli (FP) theory in de Sitter background, it is the existence of a lower bound for the mass $m$ of the spin 2 particle, the Higuchi bound. It establishes that $m^2\geq 2H^2$, where $H$ is the Hubble parameter, in order to the theory presents no ghost-like instabilities. In this sense, $m$ should be unacceptable high in order to fulfill this condition at the time of the inflationary epoch of the Universe, posing a difficulty to conciliate the FP theory with cosmology. In this article we show that the Higuchi bound can be circumvented in an alternative description of massive spin-2 particles known as $\mathcal{L}(a_1)$ models. In maximally symmetric spaces the theory has two free parameters which can be consistently chosen in order to make the model absent of a lower bound for $m$. Then, $m$ can be arbitrarily smaller than the energy scale of inflation avoiding instabilities at that time. 
\end{abstract}


\end{frontmatter}

%\linenumbers

\section{Introduction}

Since its formulation, General Relativity (GR) has successfully described gravitational interactions through a mostly geometric interpretation. The theory is well-tested over a large range of different energies and its predictions have been confirmed even today%, consolidating its theoretical foundations
. However, despite its great success, there are some theoretical impasses, such as the cosmological constant problem \cite{PCC} and the unexplained accelerated expansion of the Universe, see \cite{brax} for a recent review. According to GR, such acceleration would be caused by a ``dark energy"\, whose nature is unknown. Therefore there has been a renewal in the motivation for the search for alternative models that describe gravitation on cosmological scales and that, in addition, reproduce in certain limits the results of GR. In this sense, one of the possible extensions consists in allowing for a massive graviton.

If we assume a sufficient small mass for the graviton, it is expected that the predictions for the gravitational interaction deviate from GR only at large scales. % or equivalently late times, that is at scales comparable with the graviton Compton wave-length.
However, on flat Minkowski background, we come across the so called vDVZ discontinuity \cite{vdvz1,vdvz2}, where the nonzero mass for the graviton leads to unexpected modifications already at Solar System scales. On the other hand, the massless limit presents no problem when a cosmological constant is added to the theory. Therefore, the vDVZ discontinuity is not present in (Anti-)de Sitter spaces \cite{ref78}. %due to the existence of two mass scales – the cosmological constant H 2 and the graviton mass m 2 . 

The maximally symmetric spaces, such as de Sitter spaces, are of great interest in cosmology since we have an explicit relation between the graviton mass and the curvature of the space-time which can lead us to self-accelerating solutions for the Universe expansion. Some of the known self-accelerating cosmological solutions \cite{Deff,Deff2} consider, in a linearized approach, a massive graviton propagating in a de Sitter background. Additionally, one of the paradigms of modern cosmology is the inflationary model which predicts that the Universe went through a quasi-de Sitter expansion phase in a primitive era, prior to the era of radiation \cite{linde,linde2,Mag}. Thus, it is very relevant to study what would be the consequences for inflation if the graviton had a nonzero mass.

Although the vDVZ discontinuity is not present in the de Sitter space, a new pathological regime appears: the graviton mass must obey the inequality $m^2\geq 2H^2$, known as the Higuchi bound \cite{higuchi}, otherwise the theory acquires a ghost-like instability. More specifically, for $m^2 < 2H^2$ the helicity-0 mode of the massive graviton on the de Sitter background becomes a ghost, eliminating the possibility of studying the whole cosmological evolution starting from the high energy scales of inflation. This poses a difficulty to conciliate {models of gravity with $m\neq 0$} with cosmology.

The usual description for a massive graviton in the literature is commonly based on the Fierz-Pauli (FP) Lagrangian \cite{FP} for massive spin-2 fields. However, when terms of self-interaction are added to the Lagrangian, a non-physical degree of freedom comes up, the so called ``Boulware-Deser ghost'' \cite{bdghost}. Only in 2010, this problem could be solved in a certain decoupling limit \cite{derham1} by adding higher order graviton self-interactions with appropriately tuned coefficients and, latter, also for the complete theory, which is known as dRGT model (or ``massive gravity'') \cite{derham3}. Theories that attempt to add mass to the graviton have a long and rich history. For more details, see the review papers \cite{hinter1,derham2}. 

The study of the massive gravitons on de Sitter spaces and the Higuchi bound has been the subject of many works, such as \cite{ref76,gabad2,hig1,hig2,derham4}, which aim to analyze the implications of this restriction or even trying to circumvent it. Usually this analysis is based on the FP model in the linearized level or the non-linear bimetric theories, which are equivalent to the FP action when expanded around de Sitter metric. The question that arises is if this bound would be modified if another alternative spin-2 model is considered.

In this sense, there is a family of Lagrangians, namely $\mathcal{L}(a_1)$ \cite{Denis1,Denis2,Denis3}, where $a_1$ is a free parameter, which describes consistently massive spin-2 particles even differing essentially from the FP formulation. Additionally, for a specific value of $a_1$, we are led to a ghost-free model that does not contain the paradigmatic FP tuning, by violating the usual idea that it would be the only acceptable mass term. The coupling of these models with the gravitational background was analyzed in \cite{dalmazi2017,Hemily2} and the results point to a consistent description. The main purpose of this article is to carry this analysis further by identifying the existence (or the absence) of the Higuchi bound in the $\mathcal{L}(a_1)$ model.
Throughout the text we have used the metric signature (-,+,+,+).

\section{Higuchi bound in the FP model}

Let us start by showing the emergence  of the Higuchi bound in the FP theory. The four-dimensional FP Lagrangian considering Einstein spaces for the background curvature is given by
\begin{align} \label{FP Lagrangian}
    \mathcal{L}_{\rm FP} = & -\frac{1}{2}\nabla_\alpha h_{\mu\nu} \nabla^\alpha h^{\mu\nu} + \nabla_\alpha h_{\mu\nu}\nabla^\nu h^{\mu\alpha} \nonumber \\
    & - \nabla_\mu h\nabla_\nu h^{\mu\nu}  + \frac{1}{2} \nabla_\mu h \nabla^\mu h - \frac{1}{2}m^2(h_{\mu\nu}h^{\mu\nu} - h^2) \nonumber \\
    &+ \frac{R}{4}\Big(h^{\mu\nu}h_{\mu\nu} - \frac{1}{2}h^2 \Big), 
\end{align}
where $h_{\mu\nu}$ is a symmetric tensor, and the covariant derivatives are calculated with respect to a background metric $g_{\mu\nu}$. When dealing with the flat space, $g_{\mu\nu} = \eta_{\mu\nu}$, the Lagrangian (\ref{FP Lagrangian}) describes a ghost-free theory of massive spin-2 particles. On the other hand, in a curved background, the ghosts issue is more subtle and it depends essentially on a relation between $m^2$ and the background curvature. To proceed, let us consider the de Sitter background for which $R = 12 H^2$ is constant, where $H$ is the Hubble parameter. Now we will explicit the Higuchi bound {by using the ``cosmological'' decomposition \cite{cd1,cd2}} of $h_{\mu\nu}$ as follows
{\begin{equation}\label{h decomposition}
h_{\mu\nu} = h^{TT}_{\mu\nu} +\nabla_\mu V_\nu ^T+\nabla_\nu V_\mu^T+ g_{\mu\nu} \sigma + \nabla_\mu \nabla_\nu \tau,
\end{equation}
where $h_{\mu\nu}^{TT}$ is a transverse-traceless tensor, $V_\mu^T$ is a transverse vector, $g_{\mu\nu}$ is the de Sitter metric henceforth and the fields $\sigma$ and $\tau$ are helicity-0 degrees of freedom, unlike in Einstein's gravity where the scalar field is gauge removable.} Using Eq. (\ref{h decomposition}) in the Lagrangian (\ref{FP Lagrangian}), we find
\begin{align}
\label{lag FPs0}
\mathcal{L}_{ \rm FP,s} = & \frac{3}{2}(\nabla_\alpha \sigma)^2+\frac{3m^2}{2}\sigma  \square  \tau + \frac{3m^2H^2}{4} (\nabla_\alpha \tau)^2 \nonumber \\
&+3 (m^2-2H^2) \sigma^2,
\end{align}
where the subscript `s' means that we have written only the contribution of the scalars $\sigma$ and $\tau$ to the full FP Lagrangian. {In what follows it is enough to focus on this piece of the complete Lagrangian, since the ghost issue appears in the scalar sector of the theory.}

First of all, notice that {(for $m, H\neq0$)} the kinetic terms of both fields have the wrong sign (which should be minus in our signature), {leading to unbounded negative energy solutions} \cite{gabad3,bard}. Hence if $\tau$ had no mixing term with $\sigma$, it would be a ghost, {since we can use a kinetic mixing of two ghosts in order to eliminate one of them \cite{ref76,gabad2}}. In what follows we show how the ghosts can be eliminated from the theory for some specific values of $m^2$ {by using the same approach developed in \cite{ref76}.}

Now, in order to obtain a new Lagrangian with no mixing term, one can diagonalize the $\sigma - \tau$ kinetic terms by the shift $\sigma = \overline{\sigma} + (m^2/2)\tau$. The resulting Lagrangian is
\begin{align}\label{lag FPs}
\mathcal{L}_{ \rm FP,s} = & \frac{3}{2}(\nabla_\alpha \overline{\sigma})^2 - \frac{3m^2}{8}(m^2 - 2H^2) (\nabla_\alpha \tau)^2 \nonumber \\
&+3 (m^2-2H^2) \Big(\overline{\sigma}^2 + m^2 \overline{\sigma} \tau + \frac{m^4}{4} \tau^2\Big).
\end{align}

As one can see, the kinetic term of $\tau$ now acquires the correct sign if $m^2 > 2H^2$, while $\overline{\sigma}$ is still a ghost. 

{The equations of motion for the $\overline{\sigma}$ and $\tau$ obtained from (\ref{lag FPs}) are given by:
\begin{eqnarray}
\square \overline{\sigma}-(2H^2-m^2)(2\overline{\sigma}+m^2\tau)=0\label{eomsigma}\\
(2H^2-m^2)(\square \tau-2m^2\tau-4\overline{\sigma})=0\label{eomtau}
\end{eqnarray}
However, as we are going to see below, $\overline{\sigma}$ is not an independent dynamical field. 

 By calculating the equations of motion from the full Lagrangian, $E_{\mu\nu}=\dfrac{\delta S}{\delta h^{\mu\nu}}$, and their divergence, $\nabla^\mu E_{\mu\nu}=0$, the following constraint can be found:}
\begin{equation}
    \nabla^\mu \nabla^\nu h_{\mu\nu} = \square h,\label{constraint}
\end{equation}
and after using Eq. (\ref{h decomposition}) together with the de Sitter background we have
\begin{equation}\label{relation between the scalars}
    \square \sigma = H^2\square \tau
\end{equation}
or, in terms of $\overline{\sigma}$,
\begin{equation}\label{relation between the scalars2}
    \square \overline{\sigma} =\frac{1}{2} (2H^2-m^2)\square \tau.
\end{equation}

{Using the above constraint, the equations of motion (\ref{eomsigma}) and (\ref{eomtau}) become identical,}
\begin{eqnarray}
(2H^2-m^2)(\square \tau -2m^2 \tau -4\overline{\sigma})=0 \ ,\label{eom tau}
\end{eqnarray}
excluding $\overline{\sigma}$ from the counting of the physical degrees of freedom (with $m^2\neq 2H^2$), since it is left non-dynamical. The special case $m^2=2H^2$ will be commented just below. Thus, for $m^2 > 2H^2$, we are left with only one non-ghost helicity-0 state represented by $\tau$. This is the well known Higuchi bound. {It is worth to mention that the constraint (\ref{relation between the scalars2}) was obtained on-shell, i.e., directly from the equations of motion. For this reason, we are not allowed to use it in the Lagrangian itself \cite{moto}.} 

Note that, for exactly $m^2 = 2H^2$, the constraint (\ref{relation between the scalars2}) becomes $\square \overline{\sigma}=0$, the kinetic term of $\tau$ disappears and the equation (\ref{eom tau}) vanishes identically. On the other hand, the Lagrangian acquires a symmetry, namely,
\begin{eqnarray}
\delta h_{\mu\nu}=\nabla_\mu \nabla_\nu \phi +\frac{m^2}{2}g_{\mu\nu}\phi\ ,
\end{eqnarray}
where $\phi$ is a gauge parameter. This is the so called {\it partially massless} case \cite{higuchi,pm2,pm3,pm4,pm5,pm6,Hinter2,Hinter3}, where the scalar no longer propagates due the symmetry and the final theory has four degrees of freedom.

Finally, if $0<m^2 < 2H^2$ we have two ghost-like kinetic terms in the Lagrangian and a constraint relating them. 

By considering in (\ref{lag FPs0}) the flat massless limit ($m=H=0$), the linearized diffeomorphism symmetry of GR, $\delta h_{\mu\nu}=\partial_\mu\xi_\nu +\partial_\nu\xi_\mu$, is restored for the full Lagrangian and the scalars $\sigma$ and $\tau$ can be made zero by an appropriate gauge fixing, resulting in the absence of ghosts in the linearized Einstein's equations on flat space. 

The same procedure presented here for the FP case will be applied for the alternative spin-2 models $\mathcal{L}(a_1)$ on the next section.

\section{Alternative massive spin-2 models $\mathcal{L}(a_1)$}

The generalization of the $\mathcal{L}(a_1)$ models for curved background metrics was carried out in \cite{dalmazi2017}. The Lagrangian description of spin-2 particles in this theory employs a nonsymmetric rank-2 tensor $e_{\mu\nu} \neq e_{\nu\mu}$ which obeys the FP conditions 
\begin{equation}
    e_{[\mu\nu]} = 0, \,\,\, \nabla^\mu e_{\mu\nu}=0, \,\,\, g^{\mu\nu}e_{\mu\nu}=e=0\,\, ,
\end{equation}
in such a way that we are left with only five degrees of freedom of $e_{\mu\nu}$ in four dimensions. Focusing in the maximally symmetric spaces (MSS), the Lagrangian of the $\mathcal{L}(a_1)$ models is given by \cite{dalmazi2017}
\begin{align}\label{la1 lagrangian}
    \mathcal{L}^{\rm (MSS)}(a_1) =& -\frac{1}{4} \nabla^\mu e^{\alpha\beta} \nabla_\mu e_{\alpha\beta} - \frac{1}{4}\nabla^\mu e^{\alpha\beta}\nabla_\mu e_{\beta\alpha} \nonumber \\
    & + a_1 \nabla^\alpha e_{\alpha\beta}\nabla_\mu e^{\mu\beta} + \frac{1}{2}\nabla^\alpha e_{\alpha\beta}\nabla_\mu e^{\beta\mu} \nonumber \\
    & +\frac{1}{4} \nabla^\alpha e_{\beta\alpha} \nabla_\mu e^{\beta\mu} + \Big(a_1 + \frac{1}{4}\Big)\nabla^\mu e\nabla_\mu e \nonumber \\
    & - \Big(a_1 + \frac{1}{4}\Big)\nabla^\mu e (\nabla^\alpha e_{\alpha\mu} + \nabla^\alpha e_{\mu\alpha}) \nonumber \\
    & - \frac{m^2}{2}(e_{\alpha\beta}e^{\beta\alpha} - e^2) - \frac{1}{24}R e^{\alpha\beta} e_{\alpha\beta}\nonumber \\
    & + \Big(\tilde{f}_2 + \frac{1}{12}\Big)R e^2 \nonumber \\ 
    &- \frac{1}{4}\Big(\frac{11}{12}+ a_1 + 4\tilde{f}_2\Big)R e^{\alpha\beta}e_{\beta\alpha},
\end{align}
where $\tilde{f}_2$ is a coupling constant of $e_{\mu\nu}$ with the background geometry, while the real parameter $a_1$ enters in the kinetic terms of the Lagrangian. The FP Lagrangian is obtained for $a_1 = 1/4$ and $\tilde{f}_2 = -1/8$ in which case the antisymmetric part of $e_{\mu\nu}$ is non-dynamical.
{
Here we can see that, although the action of the ${\cal L}(a_1)$ models for curved backgrounds is different from the linearized version of the Einstein-Hilbert action, the latter can be obtained for a particular choice of the parameters. Since the Fierz-Pauli action can be obtained by setting $a_1 = 1/4$ and $\Tilde{f}_2 = -1/8$, the linearized GR is obtained in the limit $m \rightarrow 0$. Therefore, {\it in the linear regime} we can say that the Fierz-Pauli theory is a particular case of the ${\cal L}(a_1)$ models, and also the linearized Einstein-Hilbert action.

}
{The symmetries of the theory has been analyzed in \cite{Denis1}-\cite{Hemily2}. In summary, in the flat space, the massless part of $\mathcal{L}(a_1)$ is invariant under 
\begin{eqnarray}
\delta e_{\mu\nu}=\partial_\nu\xi_\mu+\partial^\alpha\Lambda_{[\alpha\mu\nu]} \ ,
\label{sym1}
\end{eqnarray}
with $\Lambda_{[\alpha\mu\nu]}$ a fully antisymmetric tensor. At $a_1=1/4$ we recover the FP model, since the antisymmetric components $(e_{\mu\nu}-e_{\nu\mu})/2$ decouple due to the enlargement of the massless symmetries (\ref{sym1}) by antisymmetric shifts $\delta e_{\mu\nu}=\Lambda_{\mu\nu}=-\Lambda_{\nu\mu}$. At $a_1=-1/12$, the massless symmetries (\ref{sym1}) are augmented by Weyl transformations $\delta e_{\mu\nu}=\eta_{\mu\nu}\phi$. The physical content of the massless version of the $\mathcal{L}(a_1)$ models has been found to be: a massless spin-2 particle plus a massless scalar for $a_1<-1/12$ or $a_1>1/4$; for $a_1=-1/12$ or $a_1=1/4$, the scalar disappears and we are left with only the massless spin-2 field; for $-1/12<a_1<1/4$, the scalar becomes a ghost. These results will be useful further in order to ensure that the massive model has a consistent massless limit.

Analogously, this analysis has been done also for the curved spaces. In \cite{Hemily2}, we have found that, for maximally symmetric spaces, the massless part of $\mathcal{L}(a_1)$ with $a_1\neq -1/12$ is indeed invariant under $\delta e_{\mu\nu}=\nabla_\nu \xi_\mu +\nabla^\alpha \Lambda_{[\alpha\mu\nu]}$, which is exactly the curved space version of the symmetry for the flat case given in (\ref{sym1}). Additionaly, the massless model obtained from the requirement of the symmetries is also consistent with the massless limit of the massive $\mathcal{L}(a_1)$ model for maximally symmetric spaces with $\tilde{f}_2=-(a_1+1/4)/4$, where we have the description of massless spin-2 particles plus massless spin-0 particles, just like in the flat case. The same analysis have been done separately for $a_1=1/12$ and the results lead us to a model consistent with the description of massless spin-2 particles propagating in maximally symmetric spaces, as we have expected due to the flat case results. More details can be found in \cite{Hemily2}.

On the other hand, on curved spaces, some local symmetries may exist even in the massive case. More specifically, the action of $\mathcal{L}(a_1)$ is invariant under the transformation
\begin{eqnarray}
\delta e_{\rho\sigma}^{(1)}=\nabla_\rho\nabla_\sigma \lambda \ ,
\end{eqnarray}
where $\lambda$ is an arbitrary scalar, provided the relation between $R$ and $m^2$ below is satisfied:
\begin{eqnarray}
R=-\frac{8m^2}{1+4a_1+16\tilde{f}_2} \ .
\end{eqnarray}

In addition, there is another possible scalar symmetry 
\begin{eqnarray}
\delta e_{\rho\sigma}^{(2)}=\nabla_\rho \nabla_\sigma \lambda +\frac{R}{12}g_{\rho\sigma}\lambda
\label{sym2}
\end{eqnarray}
where $\lambda$ is an arbitrary scalar and 
\begin{eqnarray}
R=-\frac{12m^2}{1+24\tilde{f}_2} \ .
\label{Rm2}
\end{eqnarray}

The symmetries above were identified in \cite{dalmazi2017} and, later, we could relate (\ref{sym2}) to the existence of the called partially massless theories in \cite{Hemily2}, which usually arise when the space is maximally symmetric. Moreover, the value for $R$ given in (\ref{Rm2}) is exactly the one which will lead to the lower bound of the $m^2$, regarding the Higuchi limit, as expected.}

Now, let us split the spin-2 field as {$e_{\mu\nu} = {e}_{[\mu\nu]} + e^{TT}_{\mu\nu} +\nabla_\mu V_\nu ^T+\nabla_\nu V_\mu^T  + g_{\mu\nu} \sigma + \nabla_\mu \nabla_\nu \tau$}, where again the fields $\sigma$ and $\tau$ encompasses the helicity-0 state of the massive spin-2 particle. The tensor ${e}_{[\mu\nu]}$ is the antisymmetric part of $e_{\mu\nu}$, and the symmetric part were decomposed in the same way as given by the Eq. (\ref{h decomposition}).

Although the Lagrangian $\mathcal{L}(a_1)$ is different from the FP one, from the divergence of the equations of motion, the same constraint (\ref{constraint}) can be obtained also for $\mathcal{L}(a_1)$ models. Again, the constraint can be used in the equations of motion, but not directly in the Lagrangian. As a result, one of the scalar fields can be made non-dynamical as before, and hence excluded from the counting of the physical degrees of freedom. Therefore, the `scalar sector' of the Lagrangian in (\ref{la1 lagrangian}) is 
\begin{align}\label{Lag La1 scalar}
    \mathcal{L}_{\rm s}(a_1) = & \frac{3}{8} (1 + 12a_1 )(\nabla_\alpha \sigma)^2 + 3 F(a_1,\tilde{f}_2) \sigma \, \square  \tau  \nonumber \\
    & + \frac{3}{2}H^2 F(a_1,\tilde{f}_2) (\nabla_\alpha \tau)^2 \nonumber \\
    & + \biggl[3H^2 (1 + 24\tilde{f}_2 ) + 3m^2\biggl] \sigma^2,
\end{align}
where for abbreviation we have defined
\begin{equation}
    F(a_1,\tilde{f}_2) \equiv \frac{m^2}{2} + 12H^2\biggl(\frac{a_1}{4}+\tilde{f}_2 + \frac{1}{16} \biggl).\label{Fa1f2}
\end{equation}

Notice that the above Lagrangian is a generalization of the Lagrangian (\ref{lag FPs0}), which can be easily  recovered for $a_1 = 1/4$ and $\tilde{f}_2 = -1/8$. In this specific case, the above parameter reads $F(\frac{1}{4},-\frac{1}{8}) = m^2/2$. Now, considering $a_1 \neq -1/12$, we can show that the diagonalization of the kinetic terms of the scalar fields can be carried out by using the following transformation
\begin{equation}
    \sigma = \overline{\sigma} + \frac{F(a_1,\tilde{f}_2)}{3(a_1 + \frac{1}{12})} \tau~,~~~~~a_1 \neq -\frac{1}{12},
\end{equation}
which leads to
\begin{align}\label{new scalar lagrangian}
    \mathcal{L}_{\rm s}(a_1) = &\frac{9}{2} \biggl(a_1 + \frac{1}{12} \biggl)(\nabla_\alpha \overline{\sigma})^2  \nonumber\\
    & - \frac{3}{2}F(a_1,\tilde{f}_2)\biggl[\frac{F(a_1,\tilde{f}_2)}{3(a_1 + \frac{1}{12})} - H^2 \biggl](\nabla_\alpha \tau)^2 \nonumber \\
    & + 3[m^2 + (1+24\tilde{f}_2)H^2]\biggl[\overline{\sigma} + \frac{F(a_1,\tilde{f}_2)}{3(a_1 + \frac{1}{12})}\tau\biggl]^2.
\end{align}

\begin{figure}[t!]
\begin{center}
\begin{tikzpicture}[scale=14.0]
%eixos
\draw[ultra thick][->](-0.32,0) -- (0.27,0);
\draw[ultra thick][->](0,-0.15) -- (0,0.08);
\node at (0.285,0) {$a_1$};
\node at (0,0.1) {$\tilde{f}_2$};
%funções
\draw[blue,line width = 0.05cm, smooth, domain=-1/12:0.26] plot(\x,{-1/24});
\draw[blue,line width = 0.05cm, smooth, domain=-0.32:-1/12] plot(\x,{-1/4*(\x +1/4)});
\draw[green,-, line width = 0.04cm](-1/12,-1/24) -- (-1/12,0.07);
%pontos
\filldraw[blue] (-1/12,-1/24) circle (0.003);
\filldraw[black] (-1/4,-1/24) circle (0.003);
\filldraw[black] (1/4,-1/8) circle (0.003);
%tracejadas
\draw[dashed] (-0.09,-1/24) --(-1/4,-1/24);
\draw[dashed] (-1/4,-1/24) --(-1/4,0);
\draw[dashed] (0,-1/8) --(1/4,-1/8);
\draw[dashed] (1/4,-1/8) --(1/4,0);
%coordenadas
\node at (-1/12,-0.065) {$(-\frac{1}{12},-\frac{1}{24})$};
\node at (-0.28,-0.065) {$(-\frac{1}{4},-\frac{1}{24})$};
\node at (1/4,-0.15) {$(\frac{1}{4},-\frac{1}{8})$};
%textos
\node at (-0.28,-0.09) {\footnotesize DMG};
\node at (1/4,-0.17) {\footnotesize FP};
\node at (0.12,0.045) {\bf \footnotesize Absence of};
\node at (0.12,0.02) {\bf \footnotesize Higuchi-like bound};
\node at (-1/12,0.105) {\footnotesize nFP};
%setas
%\draw[->](0.10,-0.05) -- (0.12,-0.07);
%\draw[->](-0.28,0.02) -- (-0.255,0.035);
%\draw[->](-0.075,0.015) -- (-0.06,0.03);
%equações
\node at (0.16,-0.06) {$\tilde{f}_2=-\frac{1}{24}$};
\node at (-0.24,0.04) {{$\mathbf{\tilde{f}_2=-\frac{1}{4}(a_1+\frac{1}{4})}$}};
\node at (-0.24,0.04) {{$\mathbf{\tilde{f}_2=-\frac{1}{4}(a_1+\frac{1}{4})}$}};
%\node at (-0.185,0.04) {\mathbf{$\tilde{f}_2=-\frac{1}{4}(a_1+\frac{1}{4})$}};
\node at (-0.08,0.085) {{$\mathbf{a_1=-\frac{1}{12}}$}};
%pontilhados
\draw[blue, dotted] (-0.32, 0.07) -- (0.26, 0.07);
\draw[blue, dotted] (-0.32, 0.065) -- (0.26, 0.065);
\draw[blue, dotted] (-0.32, 0.06) -- (0.26, 0.06);
\draw[blue, dotted] (-0.32, 0.055) -- (0.26, 0.055);
\draw[blue, dotted] (-0.32, 0.05) -- (0.26, 0.05);
\draw[blue, dotted] (-0.32, 0.045) -- (0.26, 0.045);
\draw[blue, dotted] (-0.32, 0.04) -- (-0.23, 0.04);
\draw[blue, dotted] (-0.19, 0.04) -- (0.26, 0.04);
\draw[blue, dotted] (-0.32, 0.035) -- (0.26, 0.035);
\draw[blue, dotted] (-0.32, 0.03) -- (0.26, 0.03);
\draw[blue, dotted] (-0.32, 0.025) -- (0.26, 0.025);
\draw[blue, dotted] (-0.32, 0.02) -- (0.26, 0.02);
\draw[blue, dotted] (-0.31, 0.015) -- (0.26, 0.015);
\draw[blue, dotted] (-0.29, 0.01) -- (0.26, 0.01);
\draw[blue, dotted] (-0.27, 0.005) -- (0.26, 0.005);
\draw[blue, dotted] (-0.23, -0.005) -- (0.26, -0.005);
\draw[blue, dotted] (-0.21, -0.01) -- (0.26, -0.01);
\draw[blue, dotted] (-0.19, -0.015) -- (0.26, -0.015);
\draw[blue, dotted] (-0.17, -0.02) -- (0.26, -0.02);
\draw[blue, dotted] (-0.15, -0.025) -- (0.26, -0.025);
\draw[blue, dotted] (-0.13, -0.03) -- (0.26, -0.03);
\draw[blue, dotted] (-0.11, -0.035) -- (0.26, -0.035);
\end{tikzpicture}
\end{center}
\caption{\raggedright \footnotesize{In this figure we show the parameter space for the $\mathcal{L}(a_1)$ models in the case of a de Sitter background metric. Each pair $(a_1, \tilde{f}_2)$ corresponds to a massive {spin-2} model. The models are free of a Higuchi-like lower bound for the mass $m$ if the pair of parameters $(a_1, \tilde{f}_2)$ is chosen in the hatched area. If the parameters are chosen in the boundary of the region represented by the blue line we have simply $m^2 > 0$. Notice that the DMG and the FP theories are out of the hatched region, while a special case of the nFP massive theory in curved spaces is represented by the vertical green line. {In addition, we have seen from the analysis of the symmetries that, in order to ensure a ghost free massless limit, we should avoid the interval $-1/12<a_1<1/4$, where the scalar becomes a ghost when the $m=0$ limit is taken.}}}\label{fig1}
\end{figure}
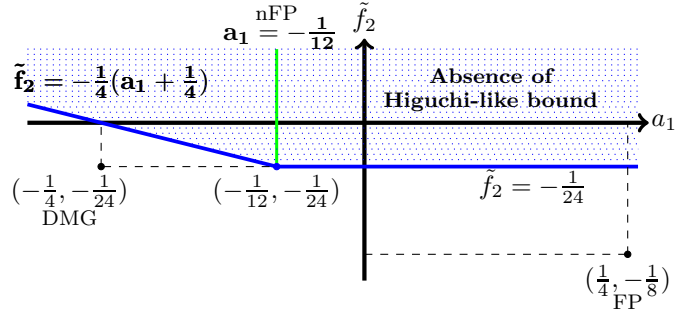

Thus, in order to obtain the correct sign for the kinetic term of the field $\tau$, there are two possibilities:
\begin{align}
    & a_1 < - \frac{1}{12} \rightarrow m^2 > - 24H^2 \biggl[\tilde{f}_2 + \frac{1}{4}\biggl( a_1 + \frac{1}{4} \biggl) \biggl], \nonumber \\
    & a_1 > - \frac{1}{12} \rightarrow m^2> -24 H^2 \biggl(\tilde{f}_2 + \frac{1}{24} \biggl).
\end{align}

Notice that, in the first case $(a_1 < -1/12)$, the field $\overline{\sigma}$ also has the correct sign for the kinetic term, while it becomes a ghost in the second case. In both cases the relation (\ref{relation between the scalars}) can be used along with the equations of motion, letting the theory with only one helicity-0 physical degree of freedom. 

From the study of the above relations, it is possible to find a region in the parameter space $(a_1, \tilde{f}_2)$ for which there is no Higuchi-like lower bound for a real mass parameter $m$. As shown in the hatched area of the Fig. \ref{fig1}, this region is delimited as follows
\begin{align}
    & a_1 < - \frac{1}{12} \rightarrow \tilde{f}_2 \geq - \frac{1}{4}\left(a_1 + \frac{1}{4}\right) , \nonumber \\
    & a_1 > - \frac{1}{12} \rightarrow \tilde{f}_2 \geq -\frac{1}{24}.
\end{align}

For any pair $(a_1, \tilde{f}_2)$ chosen in this area, the theory presents no scalar ghosts or tachyons for any real value of $m$ warranting the absence of instabilities. This is the main result of this article. On the other hand, for those values of $(a_1, \tilde{f}_2)$ out of this region, there is always a Higuchi-like lower bound for $m^2$ that is proportional to $H^2$. For instance, if $a_1 > -1/12$ and $\tilde{f}_2 = -1/8$, we find the bound $m^2 > 2 H^2$ in order to the theory presents no ghosts. The FP theory is one particular theory that enters this case. 

Another important special case located out of the above region is {\it Dual Massive Gravity} (DMG) \cite{ms} whose Lagrangian can be obtained choosing $a_1 = -1/4$ and $\tilde{f}_2 = -1/24$, which leads to the constraint $m^2 > H^2$ in order to the theory presents no ghost instabilities.

Until now we have not considered the case $a_1 = -1/12$, for which the $\mathcal{L}(a_1)$ models coincide with a special case of the non-Fierz-Pauli (nFP) massive theory \cite{DalmazinFP2015}, whose generalization to curved spaces were carried out in \cite{dalmazi2017}. In this particular case, we see from the Lagrangian (\ref{Lag La1 scalar}) that the kinetic term of the field $\sigma$ vanishes, but we have yet the kinetic term of $\tau$ and a mixing term. Considering the redefinition $\tau = \overline{\tau} + \sigma/H^2$ and, after some integration by parts, we find that scalar ghost instabilities are absent of the theory for $\tilde{f}_2 \geq -1/24$ and again we have no Higuchi-like lower bounds for $m$.

Moreover, as shown by \cite{dalmazi2017}, the above results are valid if the following restrictions are respected
\begin{equation}
\tilde{m}^2 \left[ \tilde{m}^2 + \left(a_1 -\frac{1}{4}\right)6H^2\right]\left[\tilde{m}^2 - 2H^2\right] \neq 0,
\end{equation}
\begin{equation}
\tilde{m}^2 \equiv m^2 + \left(\tilde{f}_2 + \frac{1}{8} \right)24H^2,    
\end{equation}
which imply
\begin{align}
& m^2 \neq    -24 H^2 \left(\tilde{f}_2 + \frac{1}{8}\right), \label{restriction1} \\
& m^2 \neq    -24 H^2 \left[\tilde{f}_2 + \frac{1}{4}\left(a_1 + \frac{1}{4} \right)\right], \label{restriction 2} \\
& m^2 \neq    -24H^2\left(\tilde{f}_2 + \frac{1}{24}\right). \label{restriction 3}
\end{align}

From the above restrictions, we notice that if either $\tilde{f}_2 = -1/8$, $\tilde{f}_2 = -1/24$ or $\tilde{f}_2 = -\frac{1}{4}(a_1 + \frac{1}{4})$, we have $m^2 \neq 0$. Furthermore, if the parameters $(a_1, \tilde{f}_2)$ are in the hatched region, such restrictions are trivially satisfied for a real mass. 

Finally, just like in the FP case, here we also have the partially massless models. Notice that it is possible to vanish the coefficient of $(\nabla_\alpha \tau)^2$ in (\ref{new scalar lagrangian}) by setting
\begin{eqnarray}
F(a_1,\tilde{f}_2)\biggl[\frac{F(a_1,\tilde{f}_2)}{3(a_1 + \frac{1}{12})} - H^2 \biggl]=0.
\end{eqnarray}

One possibility to satisfy this relation is if $m^2=-24H^2(\tilde{f_2} + 1/24)$, breaking the inequality (\ref{restriction 3}), since we are considering $m^2\neq0$. That was expected since, fundamentally, the partially massless models arise when, even violating the restriction (\ref{restriction 3}) for a specific relation between $m^2$ and $H$, we are still able to get all the FP conditions. See \cite{Hemily2} for details. For this specific relation between $m^2$ and $H^2$, the helicity-0 mode disappears and the theory then describes a massive spin-2 field with 4 d.o.f. in 4 dimensions. This case had already been presented for $\mathcal{L}(a_1)$ models in \cite{Hemily2}.

\begin{figure}[t!]
\begin{center}
\begin{tikzpicture}[scale=1.0]
%eixos

\draw[ultra thick][->](-0.5,0) -- (7,0);
\draw[ultra thick][->](0,-0.5) -- (0,5);
\node at (7.2,0) {$a$};
\node at (0,5.3) {$H(a)$};

%curvas
%contralada por dois pontos
\draw[line width = 0.03cm] (1,4).. controls (3, 4) and (1.5,0.5) .. (5,0.5);
%segmentos de reta
\draw[line width = 0.03cm] (-0.1,4) -- (1,4);
\draw[line width = 0.03cm] (5,0.5) -- (6.5,0.5);

%tracjadas
\draw[dashed] (-0.1,0.5) --(4.5,0.5);
\draw[dashed] (4.2,-0.1) --(4.2,4.7);

%azul
\draw[blue] (0,1) --(6.5,1);
\node[blue] at (1,1.2) {$m$};

%labels
\node at (-0.6,4) {$H_{\rm Inf}$};
\node at (-0.4,0.6) {$H^*$};
\node at (4.2,-0.3) {$a_0$};

\end{tikzpicture}
\end{center}
\caption{\raggedright \footnotesize{A sketch of the expected evolution of the Hubble function $H(a)$ in the current cosmological models is shown. For those $\mathcal{L}(a_1)$ models whose parameters $(a_1, \tilde{f}_2)$ are in the hatched area of the Fig. \ref{fig1}, it is possible to have a mass $m \ll H_{\rm Inf}$ (indicated by the blue horizontal line) avoiding scalar ghost instabilities at the time of inflation. We also indicate $H^*$, which is the asymptotic value of the Hubble function in the future if the current cosmic acceleration is driven by a cosmological constant. One should have at least $m > H_0$ in order to the graviton mass has any physical effect in the observable Universe.}}\label{fig2}
\end{figure}
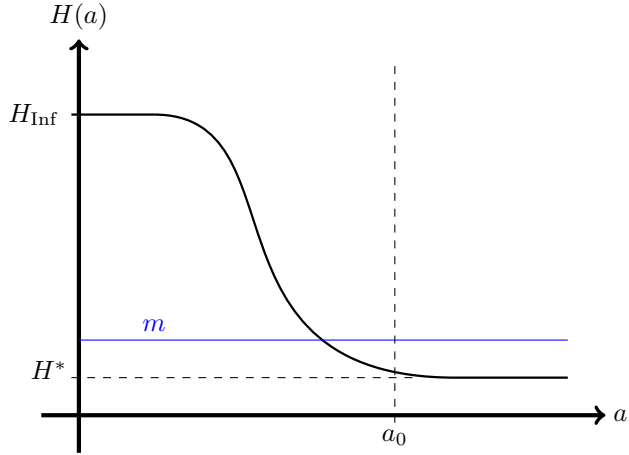

\section{Final remarks}

In a cosmological setting, the main implication of the above results is that there exists a region in the parameter space of the {massive spin-2 model} $\mathcal{L}(a_1)$ that admits a stable early de Sitter inflationary epoch, even for a graviton mass much smaller than the energy scale of inflation. The absence of scalar ghost instabilities or tachyons for any real value of the mass $m$ is warranted by choosing any pair of the parameters $(a_1, \tilde{f}_2)$ in the hatched area of the Fig. \ref{fig1}. The same is not possible for the FP and the DMG theories, for instance, which are particular cases of the $\mathcal{L}(a_1)$ model out of the hatched area. The former is free of ghosts if $m^2 > 2H_{\rm Inf}^2$ and the latter if $m^2 > H_{\rm Inf}^2$, where $H_{\rm Inf}$ is the Hubble parameter at the early de Sitter era. In both cases, $m$ should have a very high unacceptable value in order to the theory presents no ghosts at inflation. 

On the other hand, in the case of a generic $\mathcal{L}(a_1)$ model with, for instance, $a_1> -1/12$, we have $m^2 > -24H_{\rm Inf}^2 ( \tilde{f}_2 + 1/24 )$. Therefore, the right-hand-side of this inequality can be made arbitrarily small  if $\tilde{f}_2$ is arbitrarily close to $-1/24$. Moreover, the inequality is trivially satisfied for real $m$ if the right-hand-side is a negative number, i.e., $\tilde{f}_2>-1/24$. 

{This analysis also have been done for other theories such as dRGT models \cite{hig1} and bimetric gravity \cite{hig3,hig4}. The latter, especially, also seems to have less restrictive Higuchi bound, just as in the $\mathcal{L}(a_1)$ models presented here.}

{It is worth to stress that non-linear extensions of the theory for $\mathcal{L}(a_1)$ has not been studied yet, then all the present conclusions are based on the linear theory, and all the comparisons are made with the linear Fierz-Pauli action. In this sense, as far as we know, the ${\cal L}(a_1)$ models describe a massive spin-2 field as well as the Fierz-Pauli theory, with an additional freedom in the parameters that enables one to find certain combinations of them which results in the absence of the Higuchi bound.} In the near future, we expect to perform a Hamiltonian Dirac analysis of the model, which is supposed to be a nontrivial analysis since we are dealing with a nonsymmetric description for the spin-2 field. {Furthermore, in the present investigation, we have not considered the effect of introducing perturbations of the background metric. Therefore, at the moment we do not know to what extent such perturbations can change our conclusions.} 

{In view of these conclusions we see that the $\mathcal{L}(a_1)$ models have interesting features, and we are left with the issue of how to construct a full non-linear massive theory in such a way that $\mathcal{L}(a_1)$ models are obtained at the linear level of the theory.} Such a theory should accommodate a complete description of the Universe, from inflation to a future accelerated phase dominated by the cosmological constant. A sketch of the cosmological evolution of the Hubble function with the increasing of the scale factor $a$ is shown in the Fig. \ref{fig2}. Since the matter density decreases as $a^{-3}$, the only dominant component of the Universe in the future would be the cosmological constant leading to a final de Sitter epoch with the Hubble parameter $H^*$. In the past, the graviton mass could be consistently much smaller than the Hubble parameter at inflation, then with the expansion of the Universe the Hubble parameter decreases until it becomes smaller than $m$ at the present time when $H(a_0) = H_0$. This is necessary in order to the graviton mass plays any role in the observable Universe, like the acceleration of the expansion.

To conclude, it is expected that similar results could be found working with the $\mathcal{L}(a_1)$ model in a more general background geometry, as the Friedmann-Lemaître-Robertson-Walker (FLRW) metric. {Although the equations of motion change in a FLRW background, as well as the constraint equations, we expect that there is still a region in the parameter space of the theory for which we do not have a Higuchi-like bound. Also, it is expected that it is still possible to remove one of the scalars degrees of freedom.} This is a subject of further investigations.

\appendix

\section{Scalar sector of the Lagrangian}\label{app}

{
Let us start by decomposing the spin-2 field as 
\begin{eqnarray}
e_{\mu\nu} = {e}_{[\mu\nu]} + e^{TT}_{\mu\nu} +\nabla_\mu V_\nu ^T+\nabla_\nu V_\mu^T  + g_{\mu\nu} \sigma + \nabla_\mu \nabla_\nu \tau \ .\label{A1}
\end{eqnarray}

Thus, the Lagrangian given in (\ref{la1 lagrangian}), can be rewritten in terms of $\sigma$, $\tau$, $V_\mu^T$, $e_{\mu\nu}^{TT}$ and $e_{[\mu,\nu]}$.

Since we are interested only in the scalar sector of the above decomposition to analyze the Higuchi limit, we will present below only those terms of the Lagrangian containing $\sigma$ and $\tau$. Therefore, each term of the Lagrangian gives
%\begin{flalign}
\begin{eqnarray}
\nabla^\mu e^{\alpha\beta}\nabla_\mu( e_{\alpha\beta} +e_{\beta\alpha})&=& 8 (\nabla_\alpha \sigma)^2-4 \, \square \tau \, \square \sigma +2(\nabla_\alpha \nabla_\mu \nabla_\nu \tau)^2,\nonumber \\
%\nabla^\mu e^{\alpha\beta}\nabla_\mu e_{\beta\alpha} &=& 4 (\nabla_\alpha \sigma)^2-2 \, \square \tau \, \square \sigma +(\nabla_\alpha \nabla_\mu \nabla_\nu \tau)^2\nonumber \\
\nabla^\alpha e_{\alpha\beta}\nabla_\mu e^{\mu\beta} &=& (\nabla_\alpha \sigma)^2-2 \, \square \tau \, \square \sigma+(\nabla_\alpha \nabla_\mu \nabla_\nu \tau)^2+\nonumber\\
&\,& -\frac{R}{2}\sigma \, \square \tau-\frac{5}{48}R^2(\nabla_\alpha \tau)^2+\frac{R}{4}(\square \tau)^2,\nonumber \\
\nabla^\alpha e_{\alpha\beta}\nabla_\mu e^{\beta\mu} &=& (\nabla_\alpha \sigma)^2-2 \, \square \tau \, \square \sigma+(\nabla_\alpha \nabla_\mu \nabla_\nu \tau)^2+\nonumber\\
&\,& -\frac{R}{2}\sigma \, \square \tau-\frac{5}{48}R^2(\nabla_\alpha \tau)^2+\frac{R}{4}(\square \tau)^2,\nonumber \\
\nabla^\alpha e_{\beta\alpha}\nabla_\mu e^{\beta\mu} &=& (\nabla_\alpha \sigma)^2-2 \, \square \tau \, \square \sigma+(\nabla_\alpha \nabla_\mu \nabla_\nu \tau)^2+\nonumber\\
&\,& -\frac{R}{2}\sigma \, \square \tau-\frac{5}{48}R^2(\nabla_\alpha \tau)^2+\frac{R}{4}(\square \tau)^2,\nonumber \\
\nabla^\mu e\nabla_\mu e &=& 16(\nabla_\alpha \sigma)^2 - 8 \, \square \tau \, \square \sigma+(\nabla_\alpha \nabla_\mu \nabla_\nu \tau)^2+\nonumber \\
&\,&+\frac{3}{4}R (\square \tau)^2-\frac{R^2}{6}(\nabla_\alpha \tau)^2,\nonumber \\
\nabla^\mu e\nabla^\alpha( e_{\alpha\mu}+ e_{\mu\alpha}) &=& 8(\nabla_\alpha \sigma)^2-10 \, \square \tau \, \square \sigma +2(\nabla_\alpha \nabla_\mu \nabla_\nu \tau)^2+\nonumber \\
&\,&-2R \, \sigma \, \square \tau-\frac{R^2}{3}(\nabla_\alpha \tau)^2+R (\square \, \tau)^2,\nonumber \\
%e_{\alpha\beta}e^{\beta\alpha}-e^2 &=& -12\sigma^2-6\sigma \, \square \tau - \frac{R}{4}(\nabla_\alpha \tau)^2\nonumber \\
e^{\alpha\beta}e_{\alpha\beta} &=& 4\sigma^2+2\sigma \, \square \tau +(\square \, \tau)^2- \frac{R}{4}(\nabla_\alpha \tau)^2,\nonumber \\
e^{\alpha\beta}e_{\beta\alpha} &=& 4\sigma^2+2\sigma \, \square \tau +(\square \, \tau)^2- \frac{R}{4}(\nabla_\alpha \tau)^2,\nonumber \\
e^2 &=& 16\sigma^2+8\sigma \, \square \tau +(\square \, \tau)^2.
%\nabla^\alpha e_{\beta\alpha}\nabla_\mu e^{\beta\mu} &=& (\nabla_\alpha \sigma)^2-2 \, \square \tau \, \square \sigma+(\nabla_\alpha \nabla_\mu \nabla_\nu \tau)^2+\nonumber\\
%&\,& \hspace{2.4cm}-\frac{R}{2}\sigma \, \square \tau-\frac{5}{48}R^2(\nabla_\alpha \tau)^2+\frac{R}{4}(\square \tau)^2\nonumber \\
\label{scalar_terms}
\end{eqnarray}
%\end{flalign}

Notice that there are no cross-terms involving $V_\mu^T$, $e_{\mu\nu}^{TT}$ and $e_{[\mu,\nu]}$ with the scalar modes because all of them cancel out during the calculation. 

By replacing (\ref{scalar_terms}) in (\ref{la1 lagrangian}) and performing several simplifications, the scalar sector of the Lagrangian is the following
\begin{align}
    \mathcal{L}_{\rm s}(a_1) = & \frac{3}{8} (1 + 12a_1 )(\nabla_\alpha \sigma)^2 + 3 F(a_1,\tilde{f}_2) \sigma \, \square  \tau  \nonumber \\
    & + \frac{3}{2}H^2 F(a_1,\tilde{f}_2) (\nabla_\alpha \tau)^2 \nonumber \\
    & + \biggl[3H^2 (1 + 24\tilde{f}_2 ) + 3m^2\biggl] \sigma^2,
\end{align}
where we have defined
\begin{equation}
    F(a_1,\tilde{f}_2) \equiv \frac{m^2}{2} + 12H^2\biggl(\frac{a_1}{4}+\tilde{f}_2 + \frac{1}{16} \biggl),
\end{equation}
which are exactly the expressions given in (\ref{Lag La1 scalar}) and (\ref{Fa1f2}).}

\section*{Acknowledgments}

%\begin{acknowledgments}
%\noindent 

MESA would like to thank the Brazilian agency FAPESP for financial support under the thematic project \# 2013/26258-4. 

%\end{acknowledgments}

%Here are two sample references: \cite{Feynman1963118,Dirac1953888}.

%\bibliography{mybibfile}

\end{document}